\documentclass{article}

\usepackage[utf8]{inputenc} 
\usepackage[T1]{fontenc}    
\usepackage{PRIMEarxiv}
\usepackage{fancyhdr}       
\usepackage{amssymb}
\usepackage{amsfonts}       
\usepackage{amsmath}
\usepackage{amsthm}
\usepackage{amstext}
\usepackage{nicefrac}       
\usepackage{microtype}      
\usepackage[square,numbers]{natbib}
\usepackage{color}
\usepackage{hyperref}       
\usepackage{url}            
\usepackage{booktabs}       
\usepackage{multirow}
\usepackage{subcaption}
\usepackage{lipsum}
\usepackage{graphicx}       

\begin{document}

\title{Network classification through random walks}

\author{
  Gonzalo Travieso, Joao Merenda, Odemir M. Bruno \\
  Laboratory of Artificial Intelligence and Complex Systems \\ São Carlos Institute of Physics \\ University of São Paulo\\
  \texttt{joao.merenda@usp.br, gonzalo@ifsc.usp.br, bruno@ifsc.usp.br}}
  
\maketitle

\begin{abstract}
Network models have been widely used to study diverse systems and analyze their dynamic behaviors. Given the structural variability of networks, an intriguing question arises: Can we infer the type of system represented by a network based on its structure? This classification problem involves extracting relevant features from the network. Existing literature has proposed various methods that combine structural measurements and dynamical processes for feature extraction. In this study, we introduce a novel approach to characterize networks using statistics from random walks, which can be particularly informative about network properties. We present the employed statistical metrics and compare their performance on multiple datasets with other state-of-the-art feature extraction methods. Our results demonstrate that the proposed method is effective in many cases, often outperforming existing approaches, although some limitations are observed across certain datasets.
\end{abstract}

\section{Introduction}\label{sec:intro}
The study of interconnected systems through network science has advanced with applications in various disciplines, including social, biological, technological, and informational systems~\cite{costa2011applications}. By uncovering patterns and dynamic processes in complex networks, insights can be gained into the structure and functioning of the modeled systems.

The analysis of topological properties such as degree distribution, clustering coefficient, centrality measures, modularity, small-world characteristics, scale-free behavior, and hierarchical structures has furthered the understanding of emergent collective phenomena. Local patterns of interactions between individual components lead to systemic behaviors, influencing processes such as information propagation, resilience to failures, and epidemic spread~\cite{costa2007characterization}. Furthermore, modeling complex networks has enabled the prediction of behavior and the development of strategies to optimize or control dynamic processes within these interconnected systems.

The increasing availability of large-scale datasets (big data) and advances in data science and artificial intelligence have facilitated the integration of machine learning techniques into network science. This has followed two primary approaches: (i)~node classification, which involves understanding local properties and internal dynamics, and (ii)~network classification, which enables comparative analysis among distinct network structures.

Network classification requires extracting measures that can effectively characterize networks. Conventional topological and statistical metrics~\cite{costa2007characterization} are often insufficient for classification tasks, particularly in real-world networks of varying sizes. In such cases, methods capable of capturing network patterns invariant in size are necessary. To address these challenges, researchers have explored approaches based on machine learning, artificial intelligence, and automata.

Examples of successful alternatives include neural networks applied to ordered adjacency matrices~\cite{neiva2023exploring}, graph neural networks ~\cite{xin2020complex,yan2019graph,zhou2023deep}, cellular automata ~\cite{miranda2016exploring,ribas2020life,zielinski2024automata}, random walks ~\cite{backstrom2011supervised}, and deterministic walks ~\cite{silva2015high,gonccalves2012texture,gonccalves2012complex}. These methods have shown promise in characterizing network patterns and enabling effective classification.

Random walks~\cite{hughes1995} have been successfully used in diverse scientific areas, as in the study of diffusion~\cite{ibe2013elements}, modeling animal movement and foraging~\cite{maarell2002foraging}, computer vision~\cite{xia2019} and polymer physics~\cite{rubin1965random}. In network science, random walks have also been extensively studied~\cite{masuda2017rwsurvey} for applications such as search and exploration~\cite{noh2004rw,costa2007explore,fronczak2009biased}, the definition of centrality measures~\cite{noh2004rw,newman2005rwbetweenness} and network comparison~\cite{lu2014rwcompare}. Different types of networks have been considered, as trees~\cite{baronchelli2008tree}, temporal networks~\cite{perra2012temporal,starnini2012temporal}, weighted networks~\cite{zhang2013weighted,riascos2021surveyweighted}, multilayer networks~\cite{sole-ribalta2016rwmultilayer} and hypergraphs~\cite{carletti2020hypergraphs}. Furthermore, different kinds of random walkers have been employed besides the traditional random walk, as self-avoiding walks~\cite{hiley1977saw,manna1989saw,herrero2003saw,herrero2005sawscalefree,yang2005saw,lopezmillan2012saw,kim2016saw}, biased or preferential walks~\cite{costa2007explore,fronczak2009biased,bonaventura2014biased}, maximum-entropy walks~\cite{sinatra2011rwmax} and populations of interacting walkers~\cite{cencetti2018reactive}.

In this work, we use statistics of random walks with and without memory to create a set of features of networks and use these features for network classification. In Section~\ref{sec:random} we present the proposed model, in Section~\ref{sec:meth} we present the feature sets, the classification methods and the datasets used, in Section~\ref{sec:res} we present and discuss our results. Finally, our conclusions are in Section~\ref{sec:conc}.

\section{Random walks}
\label{sec:random}

Using three types of random walks: traditional, self-avoiding, and limited-memory self-avoiding walks, the characterization of networks is achieved through a set of measurements, which will be discussed below. In all cases, a walker moves from its current node to a neighboring node reachable via an outgoing link from the current node. In a \emph{traditional random walk} (RW), the next node is chosen with equal probability among all available outgoing nodes. By contrast, a \emph{self-avoiding walk} (SAW)involves keeping track of all the already visited nodes and avoiding outgoing nodes that have already been previously explored. The \emph{limited-memory} variant of the self-avoiding walk (LMW) is similar, but only retains information about the $m$ most recently visited nodes. This parameter is referred to as the memory. Regardless of the type of walk, if there are no outgoing nodes allowed, the walk ends.

For each type of walker, $W$ walkers are initiated from each node in the network. Self-avoiding walks are continued until they end; the other types of walk, which could potentially continue indefinitely or require an excessive number of steps to end, are restricted to a maximum of $S$ steps.

For each type of walk, we calculate the total number of visits to each node and normalize it by dividing it by the total number of walkers, $N W$, where $N$ is the number of nodes in the network. The starting node of the walk is not considered visited by it.  Let $t_i$ denote this normalized number of visits for node $i$ by traditional random walks, $s_i$ by self-avoiding walks, and $r^{(m)}_i$ by self-avoiding walks with memory $m$. The values of $s_i - t_i$ and $r^{(m)}_i - t_i$ serve as our starting measurements. From the distribution of these values we compute the average, standard deviation, quartiles (25\%, 50\%, and 75\% percentiles), skewness, kurtosis and entropy for each type of walk, which correspond to the features used in the classification.

In the case of the SAW, we also record the length of the walks and calculate the same statistics for it.

\section{Methodology}\label{sec:meth}

In a classification problem, the first step is to extract a set of \emph{features} of each item to be classified. The features chosen can have a strong impact on the effectiveness of the classification. In this work we compare with the random walk features proposed here four different sets of features, based on: (i)~structural network measurements, (ii)~the dynamics of life-like automata on the networks, (iii)~deterministic tourist walks, and (iv)~the \textit{graph2vec} approach. The following sections describe these methods.

\subsection{Feature extraction methods}
\label{subsec:compare}

To evaluate the performance of the proposed method, we compare it with some state-of-the-art feature extraction methods from the literature: structural measures, life-like network automata, deterministic tourist walk, and graph2vec.

\subsubsection{Structural measures}

The conventional approach to network classification involves the application of a predefined set of structural and dynamical measures to extract relevant information about the network's properties. These measures provide a comprehensive characterization of the topology of the network and its connectivity patterns, enabling a robust comparative analysis across different network types. In this study, we used the following measurements: average degree, hierarchical degree at levels 1 and 2~\cite{costa2006hierarchical}, global clustering coefficient, average shortest path length, and degree assortativity.

\subsubsection{Life-Like Network Automata}

The Life-Like Network Automata (LLNA) method is a network-based extension of cellular automata designed for pattern recognition and classification tasks. In LLNA, a network serves as the underlying structure where each vertex represents a cell with a binary state (alive or dead). The automaton evolves over discrete time steps, generating a time evolution pattern (TEP) that encodes the dynamic behavior of the system \cite{miranda2016exploring}.  

The LLNA method extends cellular automata to network-based pattern recognition and classification, where each vertex represents a binary-state cell (live or dead) evolving over discrete time steps to generate time evolution patterns (TEPs) that encode system dynamics \cite{miranda2016exploring}. Governed by life-like transition rules (Bx-Sy), which determine birth and survival based on neighborhood density, LLNA captures structural and functional network properties through its complex spatiotemporal evolution. For instance, a rule B25-S8 means that a dead cell will become alive if it has exactly 2 or 5 alive neighbors, while a living cell will survive only if it has exactly 8 alive neighbors In this paper we employed the rule \textit{B01678-S0457}.

To extract meaningful features, the LLNA-BP (Binary Pattern Descriptor) method converts TEPs into binary sequences, generating histograms that provide a robust dynamical signature without parameters~\cite{ribas2020life}.

\subsubsection{Deterministic tourist walk}

The deterministic tourist walk (DTW) method has emerged as a significant approach to texture analysis in computer vision \cite{campiteli2006deterministic,backes2010texture} and has recently been employed in network classification tasks, demonstrating high performance \cite{gonccalves2012complex}. Unlike random walks, DTW follows a deterministic movement rule based on a specific network property. In this study, we utilize the degree difference between a node and its neighbors as the guiding criterion for movement, where the tourist seeks to either minimize or maximize this difference, following the rule of minimum or maximum, respectively.

DTW is a partially self-avoiding walk, as the tourist possesses a memory of size $\mu$, preventing the revisiting of nodes stored within it. The tourist's trajectory is divided into two distinct phases:

\begin{enumerate}
    \item \textbf{Transient phase:} The tourist explores the network without getting trapped.

    \item \textbf{Attractor phase:} The tourist enters a loop and remains trapped, terminating the walk.

\end{enumerate}

Due to the memory constraint, the minimum length of an attractor must be $\mu + 1$. The method generates a unique signature for each network by constructing a trajectory histogram based on the lengths of the transient ($t$) and attractor ($p$) phases, as defined in Equation~\eqref{phi_vec}:

\begin{equation}
\label{phi_vec} 
\phi_{\mu}^{rule} = [h(\ell = \mu + 1), h(\ell = \mu + 2), \dots , h(\ell = \mu + m)] 
\end{equation}

where $\ell = t + p$ represents the total trajectory length, and $h$ is the histogram term associated with all combinations of $t$ and $p$ that result in $\ell$. The parameter $m$ is an arbitrary integer, typically set to $m=5$.

This signature vector provides an effective means of network characterization and classification.

Numerous studies have demonstrated that concatenating multiple memory sizes and movement rules significantly enhances the performance of the DTW method \cite{ribas2016fast,merenda2023using}. This approach constructs a single feature vector, defined as:

\begin{equation} \label{psi_vec} \Psi_{[\mu_1,\mu_2,\dots , \mu_n]}^{[min,max]} = [\phi_{\mu_1}^{min}, \dots , \phi_{\mu_n}^{min}, \phi_{\mu_1}^{max}, \dots, \phi_{\mu_n}^{max}] \end{equation}

In this study, we employ a combination of two memory sizes ($\mu=1$ and $\mu=2$) and two movement rules ($r=\mathrm{min}$ and $r=\mathrm{max}$), yielding the signature vector:

\begin{equation}
    \label{psi_vec2}
    \Psi_{1,2}^{min,max} = [\phi_1^{min},\phi_2^{min},\phi_1^{max},\phi_2^{max}]
\end{equation}

This representation captures diverse trajectory patterns, improving the discriminative power of the method.

\subsubsection{Graph2vec}

Graph2Vec is a method for learning graph-level embeddings, aiming to produce fixed-length vector representations of entire graphs, much like how Word2Vec generates embeddings for individual words \cite{narayanan2017graph2vec}. In contrast to conventional Graph Neural Networks (GNNs), which primarily generate embeddings for individual nodes by aggregating information from their neighbors, Graph2Vec focuses on learning whole-graph representations by identifying both structural and attribute-based similarities across graphs. This technique utilizes the Weisfeiler-Lehman (WL) subtree kernel \cite{shervashidze2011weisfeiler} to break down graphs hierarchically into subcomponents that serve as the foundation for learning embeddings.

The WL algorithm enhances node labels in an iterative fashion by hashing the labels of neighboring nodes along with each node’s current label, thereby creating progressively more detailed graph representations. During each iteration $t$, a node receives a new label derived from its current label and the set of labels from its adjacent nodes. This iterative labeling process yields a layered representation of substructures, such as rooted subtrees of depth $t$, that reflect increasingly global patterns in the graph.

Graph2Vec builds on the idea of viewing entire graphs as analogous to documents, and their constituent substructures (like neighborhoods) as analogous to words. Using a strategy similar to the skip-gram model, it learns to embed graphs in a continuous vector space such that graphs with similar structures are positioned closer together.

\subsection{Classification and validation tasks}

To classify the networks based on the feature vectors extracted using the method discussed above, we employed linear discriminant analysis (LDA). To assess the reliability and generalization of the classification results, we utilized the k-fold cross-validation strategy.

Linear Discriminant Analysis (LDA) is a supervised classification method used to find a linear combination of features that best separates two or more classes \cite{xanthopoulos2013linear}. It is particularly effective when dealing with high-dimensional data, as it reduces dimensionality while preserving class-discriminative information. The core idea behind LDA is to maximize the separation between different classes by projecting the data onto a lower-dimensional space where class means are as far apart as possible while minimizing variance within each class. This makes LDA a widely used technique in pattern recognition, machine learning, and statistical classification problems.

The LDA algorithm works by computing two key statistical measures: the between-class scatter matrix, which measures the spread of class means, and the within-class scatter matrix, which quantifies the variance of each class. The method then finds a transformation that maximizes the ratio of between-class variance to within-class variance, ensuring that projected classes remain well-separated.

To evaluate the performance of LDA, k-fold Cross-Validation, with $k=10$, was used as a validation technique. In 10-fold cross-validation, the dataset is randomly partitioned into 10 equally sized subsets, or folds. In each iteration, one fold is used as the test set, while the remaining nine folds are used for training. This process repeats 10 times, ensuring that each sample is used for testing exactly once. The final classification performance is computed as the average accuracy over all iterations \cite{wong2015performance}.

\subsection{Datasets}\label{sec:datasets}

 The datasets analyzed in this study are categorized into four distinct types: synthetic, metabolic, bioinformatics, and social.

 \begin{enumerate}
     \item \textbf{Synthetic:} This dataset is composed of two sets of synthetic networks. The first is the \textit{4-models dataset}, designed to benchmark the proposed method. The second is the \textit{noisy dataset}, constructed to evaluate the robustness of the method under varying levels of noise.
     
     \begin{enumerate}
     \item \textbf{Synthetic 4-models:} The synthetic dataset comprises 1,120 networks generated from four well-known synthetic graph models: \textit{Barabási-Albert}, \textit{Erdős-Rényi}, \textit{Watts-Strogatz}, and \textit{Waxman}. Each model includes networks of sizes 500, 1000, 1500, and 2000 nodes, with an average degree varying across 4, 6, 8, 10, 12, 14, and 16. Each class contains 280 networks.

    \item \textbf{Noisy-Dataset:} Noise was introduced into the networks from the \textbf{Synthetic 4-models} dataset using the \textbf{Link Change} (LC) technique. In this method, existing edges are randomly removed and replaced with new, randomly added edges. The extent of these modifications is controlled by the parameter $p$, referred to as the \textit{Noise rate}, which ranges from $0\%$ to $100\%$. This parameter controls the extent of modifications applied to the network. Specifically, for a given rate $p'$, half of this proportion ($p'/2$) of the existing edges are removed, and the other half ($p'/2$ of the original number of edges) are added.
    To construct the noisy dataset, we employed ten discrete values of $p$: 10, 20, 30, 40, 50, 60, 70, 80, 90, and 100.
    
 \end{enumerate}

 \item \textbf{Metabolic:} The metabolic dataset was obtained from the \textit{Kyoto Encyclopedia of Genes and Genomes} (KEGG) database \cite{kanehisa2000kegg,kanehisa2016kegg}. It consists of seven sets of metabolic networks, where each node represents a metabolite, and edges correspond to biochemical reactions linking metabolites from substrates to those in products.

 \begin{enumerate}
     \item \textbf{Kingdom-database:} includes species from the eukaryota domain, covering four kingdoms; \textit{animals}, \textit{plants}, \textit{fungi}, and \textit{protists}, each containing 40 networks.

     \item \textbf{Animal-database:} comprises four classes; \textit{mammals}, \textit{birds}, \textit{fishes}, and \textit{insects}, with 14 samples per class.

     \item \textbf{Fungi-database:} consists of four fungal classes; \textit{saccharomycetes}, \textit{sordariomycetes}, \textit{eurotiomycetes}, and \textit{basidiomycetes}, each containing 15 networks.

      \item \textbf{Plant-database:} includes three categories; \textit{monocots}, \textit{green algae}, and \textit{eudicots}, each with nine organisms.

    \item \textbf{Protist-database:} contains four groups; \textit{Amoebozoa}, \textit{Alveolates}, \textit{Stramenopiles}, and \textit{Euglenozoa}, with five organisms in each.

     \item \textbf{Firmicutes-Bacillis-database:} consists of four classes; \textit{Bacillus}, \textit{Staphylococcus}, \textit{Streptococcus}, and \textit{Lactobacillus}, with an unbalanced number of species: 122, 76, 133, and 83, respectively.

    \item \textbf{Actinobacteria-database:} also presents an uneven number of species among its three classes; \textit{Mycobacterium}, \textit{Corynebacterium}, and \textit{Streptomyces}, containing 60, 86, and 53 species, respectively.
 \end{enumerate}

 \item \textbf{Bioinformatics:} This category comprises two datasets, \textit{Enzymes} and \textit{Proteins}, both sourced from the \textit{Technischen Universität Dortmund Dataset} (\textit{TUDataset}) \cite{Morris+2020}. The \textit{Enzymes} dataset contains 600 networks, evenly distributed across six classes. In contrast, the \textit{Proteins} dataset consists of 1,113 networks, divided into two classes containing 663 and 450 networks, respectively.

 \item \textbf{Social:} This collection comprises three datasets extracted from the \textit{TUDataset} \cite{Morris+2020}. 

 \begin{enumerate}
     \item \textbf{Collab:} This dataset consists of collaboration networks, where each node represents a researcher and edges denote collaborative relationships. It includes $5,000$ networks categorized into three classes, containing 2,600, 775, and 1,625 networks, respectively.

     \item \textbf{IMDB-Multi:} A multiclass social network dataset consisting of 1,500 networks evenly distributed across three distinct classes, with 500 networks in each class.

 \end{enumerate}

 \end{enumerate}

 \subsection{Experimental setup}

For each of the different types of random walks described in Section~\ref{sec:random}, $W=10$ walkers starting on each node were done. With the exception of the self-avoiding walks, the walkers were limited to $S=N$ steps, where $N$ is the number of nodes in the network. Memory-limited walkers were done for memory sizes of $m=1\ldots10$.

Each of the measurement described in Section~\ref{sec:random} were computed and their combinations were used as features to a LDA classifier to perform the classification of the networks. The classification performance of the various combinations of measurements were analyzed and then compared against several baseline and state-of-the-art methods, as detailed in Section~\ref{subsec:compare}.
 
\section{Results}\label{sec:res}

First, we evaluated the performance of each individual feature set and their combinations. Tables~\ref{tab:1}, \ref{tab:2}, and \ref{tab:3} present the classification accuracy for various combinations of feature sets derived from the Random Walk method across multiple network datasets. Among the individual sets, the one base on statistics of visited nodes by self-avoiding walks consistently achieved the highest accuracy compared to statistics of self-avoiding walk lengths, and any single size of memory-limited walk. Notably, the SAW visits set alone outperformed the others measure sets in 11 of the 12 datasets, highlighting the informativeness of node visitation statistics for network classification.

Looking at the combinations of features, the pairing of SAW lengths and visits really took the lead, achieving the best results in 8 out of 12 datasets. Meanwhile, the broader combination with all features achieved the best performance in 3 datasets but offered only marginal gains (typically less than $1$ - $2\%$) over the features from taken from the SAW. Given the significant computational demands of handling and storing the complete memory-based feature set, the restriction to the use of only self-avoiding walks (together with the baseline traditional random walks) offers a great compromise between performance and efficiency.

\begin{figure*}[h!]
    \centering
    \includegraphics[scale=0.5]{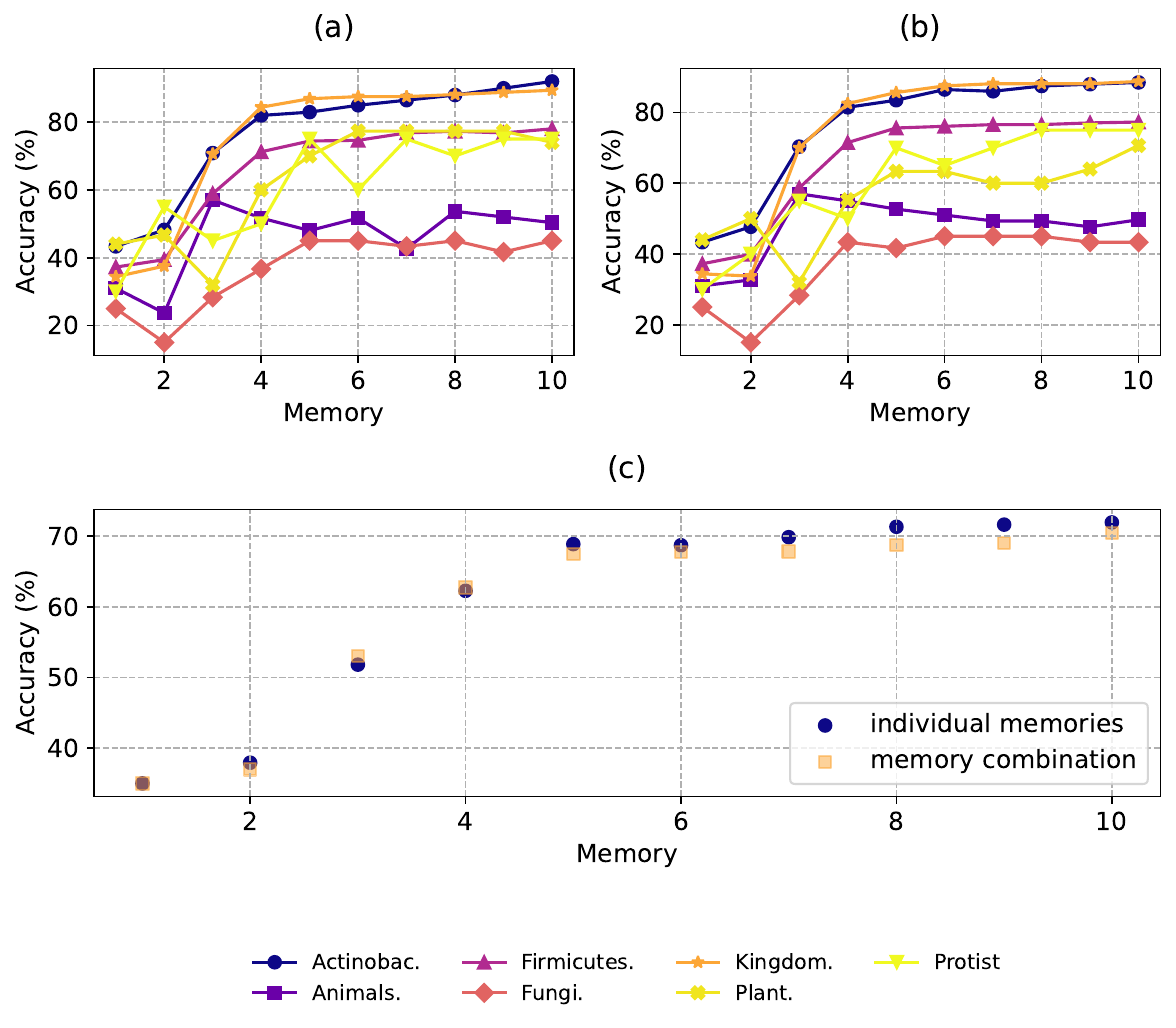} 
    \caption{Classification performance of the limited-memory random walk as a function of the memory parameter $m$ in the metabolic datasets.
(a)~Accuracy obtained using individual memory values ($m = 1$ to $10$) in isolation.
(b)~Accuracy obtained using cumulative combinations of memory values, incrementally adding from $m = 1$ to $m = 10$.
(c)~Average accuracy across all metabolic datasets as the memory increases, comparing individual memory usage and memory combinations.
}
\label{fig:performance}
\end{figure*}

The analysis presented in Figure~\ref{fig:performance} further supports this conclusion. The classification accuracy initially increases with the memory value in the limited-memory random walks but quickly reaches a plateau. For most datasets, performance saturates around memory values of 7 or 8, with no meaningful improvement beyond this point. This indicates diminishing returns when incorporating deeper memory in the walk.

Table~\ref{tab:4} summarizes the comparative performance of the proposed method and several state-of-the-art baselines, including structural measurements, LLNA, DTW, and Graph2vec. The results clearly show the competitiveness of the proposed model: using only the combination of SAW statistics, our method achieves the highest accuracy in 9 out of 12 datasets. It ties in one case (\emph{Synthetic 4-models}) and is outperformed only in two cases: \emph{Firmicutes-Bacillus} and \emph{Collab}. 

\newpage

\begin{table*}[h!]
    \centering
    \begin{tabular}{lrrrr}
        \toprule
        Feature set & Actinobac. & Animals & Firmicutes & Fungi \\
        \midrule
        SAW length & 37.3 (9.8) & 76.6 (13.9) & 32.1 (1.0) & 46.6 (8.1) \\
        SAW visits       & 95.9 (5.8) & 80.0 (12.2) & 87.2 (3.2) & 63.3 (11.0)  \\
        LMW $m=1$  & 43.3 (8.8) & 31.0 (15.2) & 37.2 (6.2) & 25.0 (13.4) \\
        LMW $m=2$  & 48.2 (11.3) & 23.6 (16.0) & 39.4 (3.7) & 15.0 (11.6)  \\
        LMW $m=3$  & 70.8 (8.6) & 57.0 (11.4) & 58.8 (5.7) & 28.3 (13.0)  \\
        LMW $m=4$   & 81.9 (5.4) & 51.6 (15.2) & 71.2 (4.6) & 36.6 (14.5) \\
        LMW $m=5$  & 82.9 (4.5) & 48.0 (15.0) & 74.4 (2.6) & 45.0 (10.6)  \\
        LMW $m=6$  & 84.9 (7.3) & 51.6 (13.2) & 74.6 (3.3) & 45.0 (10.6) \\
        LMW $m=7$ & 86.4 (4.9) & 42.6 (16.0) & 76.8 (2.9) & 43.3 (15.2) \\
        LMW $m=8$ & 87.9 (7.4) & 53.6 (12.8) & 77.2 (1.9) & 45.0 (13.0) \\
        LMW $m=9$  & 89.9 (7.4) & 52.0 (12.1) & 76.7 (2.8) & 41.6 (15.3)  \\
        LMW $m=10$ & 91.9 (3.9) & 50.3 (13.3) & 78.0 (2.4) & 45.0 (13.0) \\
        \midrule
        SAW length $+$ visits   & 96.4 (5.4) & \textbf{96.4 (5.4)} & \textbf{90.2 (5.4)} & \textbf{75.1 (5.4)}  \\
        SAW length $+$ visits + LMW $m=1, 2$ &  90.9 (5.3) &  86.6 (9.1)  & 87.2 (3.2)   &  58.3 (13.0)   \\
        SAW length $+$ visits + LMW $m=1\ldots3$   &  93.9 (6.9) &  87.0 (11.4)  & 86.5 (3.2)   &   66.6 (12.4)  \\
        SAW length $+$ visits + LMW $m=1\ldots4$   &  93.4 (6.3) &  81.3 (18.4)  & 84.8 (4.2)   &  61.6 (15.3)   \\
        SAW length $+$ visits + LMW $m=1\ldots5$   &  93.9 (4.2) & 81.0 (16.8)   &  85.6 (2.7)  &   61.6 (11.1)   \\
        SAW length $+$ visits + LMW $m=1\ldots6$   &  93.9 (4.5) & 89.3 (17.6)   & 86.0 (2.5)   &  65.0 (13.0)  \\
        SAW length $+$ visits + LMW $m=1\ldots7$   &  96.4 (5.9) & 89.3 (17.6)   &  86.3 (2.6)  &  65.0 (13.0)  \\
        SAW length $+$ visits + LMW $m=1\ldots8$   &  97.4 (5.5) & 87.6 (18.2)   &  86.5 (2.7)  &  65.0 (13.0)  \\
        SAW length $+$ visits + LMW $m=1\ldots9$   &  97.9 (5.9) &  87.6 (18.2)  &  86.5 (2.7)  &   73.3 (13.3) \\
        SAW length $+$ visits + LMW $m=1\ldots10$   &  \textbf{98.4 (4.9)} &  89.6 (16.0)  &  87.2 (2.3)  &  73.3 (13.3)    \\
        \bottomrule
    \end{tabular}
    \caption{Average classification accuracy (in $\%$) and standard deviation for different feature set combinations based on the Random Walk method.}
    \label{tab:1}
\end{table*}

\begin{table*}[h!]
    \centering
    \begin{tabular}{lrrrr}
        \toprule
        Feature set & Kingdom & Plant & Protist  & Enzymes   \\
        \midrule
        SAW length & 52.5 (8.0) &55.3 (17.5) & 35.0 (33.9) &  24.0 (4.4) \\
        SAW visits   & 89.3 (4.8)   & 74.0 (9.0) & 70.0 (10.0) &  26.8 (3.9)  \\
        LMW $m=1$ & 34.3 (10.5) & 44.0 (13.5) & 30.0 (24.4) &  15.6 (5.4) \\
        LMW $m=2$  & 37.5 (8.3)  & 46.6 (17.8) & 55.0 (10.0) &  20.6 (4.7)  \\
        LMW $m=3$  & 70.6 (8.8) & 32.0 (19.3) & 45.0 (10.0) &  20.5 (3.4)   \\
        LMW $m=4$  & 84.3 (7.5) & 60.0 (15.2) & 50.0 (27.3) &  19.6 (5.5)  \\
        LMW $m=5$ & 86.8 (5.9) & 70.0 (9.8) & 75.0 (15.8) &  20.1 (4.4)  \\
        LMW $m=6$ & 87.5 (6.2) & 77.3 (13.7) & 60.0 (12.2) &  18.3 (3.3)  \\
        LMW $m=7$ &  87.5 (5.5) & 77.3 (8.7) & 75.0 (15.8) &  16.5 (4.2) \\
        LMW $m=8$ & 88.1 (5.9)  & 77.3 (8.7) & 70.0 (10.0) &  18.3 (4.0) \\
        LMW $m=9$  &  88.7 (4.6) & 77.3 (13.7) & 75.0 (15.8) &  16.3 (4.0)  \\
        LMW $m=10$ &  89.3 (4.0)  & 74.0 (9.0) & 75.0 (15.8) &  16.0 (3.0) \\
        \midrule
        SAW length $+$ visits & \textbf{98.7 (4.6)} & \textbf{88.3 (5.4)} & 80.2 (5.4)  & \textbf{27.9 (5.4)}  \\
        SAW length $+$ visits + LMW $m=1, 2$ & 89.3 (4.8)  &  90.9 (5.3) &  70.0 (10.0)  &      22.6 (3.8)    \\
        SAW length $+$ visits + LMW $m=1\ldots3$ & 80.6 (7.1)  &  39.3 (21.9) &  70.0 (10.0)  &      22.5 (3.0)  \\
        SAW length $+$ visits + LMW $m=1\ldots4$   & 84.3 (8.9)  &  59.3 (13.7) &  75.0 (25.5)  &     26.1 (2.5)   \\
        SAW length $+$ visits + LMW $m=1\ldots5$   & 86.2 (7.8)  &  63.3 (9.8) & 70.0 (18.7)   &      24.8 (3.3) \\
        SAW length $+$ visits + LMW $m=1\ldots6$  & 87.5 (7.4)   &  63.3 (9.8) & 75.0 (20.0)   &    23.8 (3.2)   \\
        SAW length $+$ visits + LMW $m=1\ldots7$ &  88.1 (5.9)  &  66.0 (15.2) & 80.0 (18.7)   &     22.6 (3.8)  \\
        SAW length $+$ visits + LMW $m=1\ldots8$  & 91.1 (5.9)   &  66.0 (15.2) & 80.0 (15.8)   &     23.0 (3.4)   \\
        SAW length $+$ visits + LMW $m=1\ldots9$ &  85.1 (5.9)  &  70.0 (17.1) &  82.0 (15.8)  &       21.8 (4.1)  \\
        SAW length $+$ visits + LMW $m=1\ldots10$ &  96.7 (4.6)  &  77.6 (8.0) &  \textbf{82.0 (15.8)}  &      26.0 (3.9)   \\
        \bottomrule
    \end{tabular}
    \caption{Average classification accuracy (in $\%$) and standard deviation for different feature set combinations based on the Random Walk method.}
    \label{tab:2}
\end{table*}

\newpage

\begin{table*}[h!]
    \centering
    \begin{tabular}{lrrrr}
        \toprule
        Measure set & Proteins & Collab & IMDB-Multi  & Synthetic  \\
        \midrule
        SAW length & 64.6 (1.3) & 40.2 (3.1) &  96.2 (1.0) & 71.2 (4.6) \\
        SAW visits       & 59.6 (1.8) & 39.0 (2.4) &  95.2 (0.6) & 70.9 (2.9) \\
        LMW $m=1$  & 54.8 (0.8) & 34.4 (1.1) &  81.3 (1.0) & 66.1 (3.1) \\
        LMW $m=2$  & 53.8 (1.0) & 36.2 (2.7) &  80.7 (0.7) & 67.5 (3.0)\\
        LMW $m=3$  & 53.6 (0.4) & 36.0 (2.5) &  91.4 (0.8)  & 65.5 (2.5) \\
        LMW $m=4$   & 53.2 (1.0) & 37.2 (2.0) &  95.0 (0.8) & 67.5 (3.0)  \\
        LMW $m=5$  & 53.2 (0.9) & 35.5 (2.2) &  96.0 (0.8) & 69.0 (3.7) \\
        LMW $m=6$  & 53.8 (0.7) & 39.8 (1.5) &  96.9 (1.1) & 68.4 (3.5) \\
        LMW $m=7$ & 53.4 (0.5) & 40.6 (2.2) &  97.7 (0.9) & 68.1 (2.5)\\
        LMW $m=8$ & 53.0 (0.5) & 39.8 (2.1) &  98.1 (0.9) & 68.8 (2.4)\\
        LMW $m=9$  & 52.8 (0.6) & 39.1 (2.6) &  98.1 (0.8) & 70.4 (2.7) \\
        LMW $m=10$ & 55.9 (2.3) & 38.9 (1.9) &  98.4 (1.0) & 69.8 (3.0)\\
        \midrule
        SAW length $+$ visits   & 65.1 (5.4) & \textbf{49.3 (5.4)}  & \textbf{100.0 (0.0)} & 78.1 (5.4) \\
        SAW length $+$ visits + LMW $m=1, 2$   &  65.1 (1.3) &  41.0 (3.1)  &      98.4 (0.9)   & 71.9 (4.0)  \\
        SAW length $+$ visits + LMW $m=1\ldots3$   &  65.7 (1.2) &  41.7 (3.3)  &      100.0 (0.0)  & 71.7 (4.0) \\
        SAW length $+$ visits + LMW $m=1\ldots4$   &  66.0 (1.2) &  41.6 (3.2)  &     100.0 (0.0)  & 71.5 (3.8)  \\
        SAW length $+$ visits + LMW $m=1\ldots5$   &  65.9 (1.2) & 40.6 (3.8)   &      100.0 (0.0) & 74.9 (3.4)  \\
        SAW length $+$ visits + LMW $m=1\ldots6$   &  65.9 (1.4) & 39.6 (3.0)   &    99.21 (1.5)  & 76.7 (3.8)  \\
        SAW length $+$ visits + LMW $m=1\ldots7$   &  \textbf{66.2 (1.2)} & 40.4 (3.3)   &     100.0 (0.0)  & 76.3 (3.9)   \\
        SAW length $+$ visits + LMW $m=1\ldots8$   &  65.7 (1.9) & 41.0 (3.0)   &      100.0 (0.0)  & 76.2 (3.8)   \\
        SAW length $+$ visits + LMW $m=1\ldots9$   &  64.8 (2.0) &  41.1 (3.1)  &       100.0 (0.0) & 77.1 (4.1)  \\
        SAW length $+$ visits + LMW $m=1\ldots10$   &  63.8 (2.0) &  41.4 (3.1)  &      100.0 (0.0) & \textbf{78.43 (3.7)}   \\
       \bottomrule
    \end{tabular}
    \caption{Average classification accuracy (in $\%$) and standard deviation for different feature set combinations based on the Random Walk method.}
    \label{tab:3}
\end{table*}

\begin{table*}[h!]
    \centering
    \begin{tabular}{lrrrrr}
        \toprule
        Dataset & Structural & LLNA & DTW & Graph2vec & Random walks \\
        \midrule
        Synthetic 4-models & $100.0$ ($0.0$) & $100.0$ ($0.0$) & $100.0$ ($0.0$) & $100.0$ ($0.0$) & $100.0$ ($0.0$)\\
        Actinobacteria       & $93.2$ ($0.7$) & $95.1$ ($1.2$) & $94.9$ ($1.4$) & $96.3$ ($15.9$) & $\mathbf{96.4}$ ($5.4$)  \\
        Animal   & $83.7$ ($15.2$) & $84.9$ ($15.2$) & $80.0$ ($17.2$) & $92.1$ ($26.8$) &$\mathbf{96.4}$ ($5.4$) \\
        Firmicutes-Bacillus  & $95.7$ ($0.6$) & $\mathbf{98.3}$ ($1.2$) & $93.3$ ($2.3$) & $98.2$ ($13.1$) & $90.2$ ($5.4$) \\
        Fungi  & $54.9$ ($15.4$) & $74.2$ ($17.4$) & $68.6$ ($14.8$) & $74.4$ ($22.4$) &$\mathbf{75.1}$ ($5.4$) \\
        Kingdom  & $96.6$ ($4.3$) & $97.4$ ($4.0$) & $89.6$ ($4.0$) & $98.4$ ($12.2$) &$\mathbf{98.7}$ ($5.6$)  \\
        Plant   & $54.2$ ($9.2$) & $74.8$ ($5.6$) & $59.9$ ($6.6$) & $75.7$ ($12.8$) & $\mathbf{88.3}$ ($5.4$) \\
        Protist  & $45.1$ ($10.9$) & $80.0$ ($5.3$) & $61.4$ ($13.7$) & $64.2$ ($17.9$) &$\mathbf{80.2}$ ($5.4$) \\
        Enzymes & $0.0$ ($0.0$) & $3.0$ ($1.7$) & $0.0$ ($0.0$) & $3.1$ ($17.5$) & $\mathbf{27.9}$ ($5.4$)\\
        Proteins & $41.1$ ($13.1$) & $59.1$ ($17.0$) & $49.0$ ($21.1$) & $64.1$ ($27.9$) & $\mathbf{78.1}$ ($5.4$)\\
        Collab  & $47.2$ ($3.7$) & $49.2$ ($4.9$) & $52.3$ ($12.1$) & $\mathbf{75.5}$ ($13.0$) & $65.1$ ($5.4$)\\
        IMDB-Multi & $14.8$ ($17.4$)& $39.7$ ($14.1$) & $33.6$ ($20.7$) & $36.9$ ($18.5$) & $\mathbf{49.3}$ ($5.4$)\\
        
        \bottomrule
    \end{tabular}
    \caption{Average $\%$ and (standard deviation $\%$) of the accuracies of the methods used in this study: structural measurements, life-like network automata (LLNA), deterministic tourist walk (DTW), Graph2vec and the random walk based approach proposed here. The best classification for each dataset is shown in bold face.}
    \label{tab:4}
\end{table*}

\begin{figure*}[h!]
    \centering
    \includegraphics[width=0.6\textwidth]{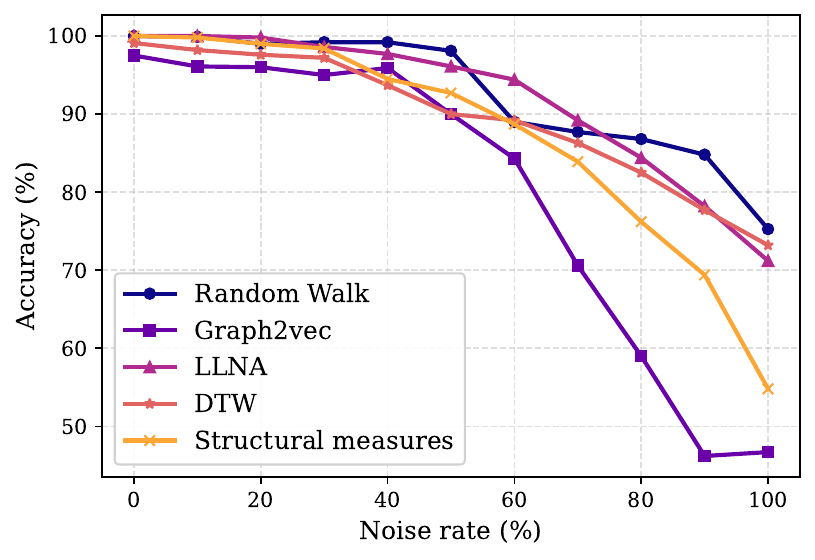}
    \caption{Classification performance under increasing noise levels. The proposed Random Walk-based method (using the \emph{sas} and \emph{sav} feature sets) is compared against state-of-the-art methods.}
    \label{fig:2}
\end{figure*}

We also want to evaluate how sensitive the classification based on random walk features is to a noise in the network dataset. For this, we make use of the noisy datasets described in Section~\ref{sec:datasets}. As illustrated in Figure~\ref{fig:2}, the random walk-based method demonstrates strong robustness to noise, achieving superior performance on the noisy dataset. The proposed method exhibited the lowest average performance drop, with a loss of only $24.73\%$. In comparison, the DTW method experienced a $26.13\%$ loss, while LLNA and structural measures suffered more substantial declines of $28.80\%$ and $45.20\%$, respectively. Notably, the Graph2vec method showed the highest sensitivity to noise, with a performance degradation of $52.10\%$. These results highlight the resilience of the proposed approach under increasing noise levels, maintaining significantly higher accuracy than competing methods across the entire noise spectrum.

These findings reinforce that the proposed method not only achieves strong overall classification performance but also exhibits robustness and efficiency. Its superior performance in most datasets, combined with its resilience to noise, makes it a promising and scalable approach for network classification tasks across diverse domains.

\section{Conclusions}
\label{sec:conc}
In this work, we presented a method for the extraction of features of networks based on different types of random walks: traditional, self-avoiding and limited memory self-avoiding walks. Statistics of node visits of all this kinds of walks and of walk lengths for self-avoiding walks are combined and used to characterize a network with the intention of network classification.

The results demonstrate that the combination of traditional random walks and self-avoiding random walks is sufficient to achieve high classification accuracy in network datasets. The inclusion of limited-memory random walks does not significantly enhance performance and can be considered unnecessary, especially when accounting for the increased computational cost. The proposed method, based on the combined node visit and walk lenght features from self-avoiding walks outperformed state-of-the-art baseline methods in 10 out of the 12 evaluated datasets, achieving perfect accuracy ($100\%$) in the \emph{Synthetic 4-models} dataset. However, in two datasets, \emph{Firmicutes-Bacillus} and \emph{Collab}, the method showed limited performance, indicating that further investigation may be needed to address specific structural properties in these cases.

In addition, the proposed approach exhibited strong robustness under noise, as evidenced by its performance on the \emph{Noisy dataset}. It showed only a $24\%$ drop in accuracy, which is substantially lower than the degradation observed in competing models, confirming the method’s resilience to perturbations in network structure. 

The performance of the proposed method could be further improved by incorporating additional dynamic features extracted from the random walks. For instance, using first visit times, which capture the step at which each node is first reached during a walk, can provide complementary temporal information that may enhance discriminative power. In particular, such enriched feature sets may help address the limitations observed in datasets where the method underperformed, such as \emph{Firmicutes-Bacillus} and \emph{Collab}.

One limitation of the proposed method is its computational cost, which increases with the size of the network. Specifically, both the number of walkers and the length of the walks scale with the number of nodes, making the approach less suitable for large-scale networks. A potential solution is to impose a fixed upper limit on the walk length, regardless of network size. However, the impact of this constraint on classification accuracy and feature quality remains an open question and should be systematically investigated in future work.

\section*{Acknowledgments}
O. M. B. acknowledges support from FAPESP (grant \#21/08325-2) and CNPq (grant \#305610/2022-8). 
J. M. acknowledges support from CAPES.

\bibliographystyle{unsrt}

\end{document}